# Stable polarization-encoded quantum key distribution in fiber

Guang Wu, Jie Chen, Yao Li, and Heping Zeng[1]

Key Laboratory of Optical and Magnetic Resonance Spectroscopy, and Department of Physics, East China Normal University, Shanghai 200062, People's Republic of China

**Abstract**

Polarizations of single-photon pulses have been controlled with long-term stability of more than 10 hours by using an active feedback technique for auto-compensation of unpredictable polarization scrambling in long-distance fiber. Experimental tests of long-term operations in 50, 75 and 100 km fibers demonstrated that such a single-photon polarization control supported stable polarization encoding in long-distance fibers to facilitate stable "one-way" fiber system for polarization-encoded quantum key distribution, providing quantum bit error rates below the absolute security threshold.

---

[1] Corresponding authors. Electronic address: hpzeng@phy.ecnu.edu.cn.

Quantum key distribution (QKD) has nowadays been demonstrated as a cryptographic approach to provide absolute security between transmitter and receiver [1,2]. Fiber-based systems have been demonstrated in prototype QKD implementations, with practical stabilities in long-distance telecom fibers [3-6]. Among those, the so-called "plug and play" QKD system realized in "two-way" fibers is to date the most stable one due to its phase-drift balance of Mach-Zehnder (MZ) interferometer arm lengths and automatic compensation of any birefringence effects and polarization-dependent losses in the transmission fibers [3,4]. However, the "two-way" QKD may be threatened by Trojan attacks and is difficult to implement with single-photon sources. A "one-way" QKD can be realized with phase encoding in asymmetric interferometers at the transmitter and receiver sites [5]. Although influence of polarization fluctuations in long-distance transmission fibers on single-photon interference can be overcome by using Faraday reflection in modified Michelson interferometers [6], interference stability and thus quantum bit error critically depend on phase shifts. A continuous active control of the interferometer arm lengths is required for these phase-encoded fiber QKD systems. With an appropriate feedback control, phase-drift errors can nowadays readily be reduced low enough for practical systems. On the other hand, polarization encoding can be also used for fiber QKD, which can be implemented without any active modulation elements at Alice's or Bob's sites or in the quantum channel (the fiber connection between Alice and Bob) [7-10]. As compared with a phase-encoded fiber system, polarization-encoded QKD uses less lossy elements. The key-creation efficiency can thus be possibly increased with an improved security in practice. Nevertheless, polarization-encoded "one-way" fiber systems are difficult to make practical, due to the unpredictable polarization scrambling imposed by randomly induced birefringence in installed telecom fibers or polarization-dependent losses of on-line optical elements within the systems.

In this letter, we report on a feedback control of single-photon polarization that facilitates active long-term polarization stabilization to beat the unpredictable polarization scrambling in long-distance fibers. With single-photon polarizations being actively stabilized, stable polarization-encoding can be readily realized in fiber systems for efficient "one-way" QKD. In the experimental implementation, commercially available polarization-controlling elements typically have very low insertion losses and can be easily installed on-line in fiber systems. We demonstrated an active compensation of randomly induced fiber birefringence in a "one-way" polarization-encoded QKD fiber system with a distance up to 100 km and long-term stability over 10 hours. Polarization feedback control at single-photon level can also be used as a very robust part in a "one-way" phase-encoded fiber system since it can simplify the necessary polarization control by replacing lossy passive elements with simple active ones.

After transmission of long-distance fibers, a linearly polarized beam at Alice's site changes to be randomly polarized at Bob's site, mainly due to birefringence induced by the unavoidable stress or asymmetry of fiber. A linear polarization along the direction $+45^o$, $-45^o$, $0^o$ or $90^o$ at Alice's site, represented respectively by the point Q, R, H or V in the equator of the Poincare sphere as shown in the inset of Fig .1(a), may change into an arbitrary state of polarization (SOP) at Bob's site, represented for instance by the point P on the Poincare sphere. In order to have accurate polarization decoding in the presence of unpredictable polarization scrambling, Bob needs a polarization feedback control to rotate his random SOPs back to the original ones sent by Alice. This can be realized with electronic polarization controllers consisting of properly aligned piezoelectronic actuators to stress on fiber along different directions, which induce desired birefringence in fiber to adjust the SOP in a controllable way [11-14]. In the experiment, we used two electronically controlled piezoelectronic actuators $X_1$ and $X_2$ to stress the fiber along $45^o$ and $0^o$, respectively. According to the visualized representation of polarizations in the Poincare sphere [see the inset of Fig. 1(a)], an applied piezoelectronic voltage on $X_1$ brings about a clockwise rotation of the SOP along the QR axis, corresponding to a phase retarder between the eigenmodes (i.e., linear polarizations along $\pm 45^o$) of the induced birefringence, while an applied piezoelectronic voltage on $X_2$ causes the SOP to rotate along the HV axis clockwise, corresponding to a phase retarder between the eigenmodes polarizations (along $0^o$ and $90^o$) of the induced birefringence. A proper combination of piezoelectronic voltages on $X_1$ and $X_2$ can compensate for the arbitrary SOP changes of orthogonal polarizations, corresponding to rotations along QR and HV axes in a proper series. As only orthogonal SOPs cover the same rotations along the HV and QR axes, this is inapplicable to simultaneous control of nonorthogonal SOPs.

Polarization-encoded QKD involves at least two nonorthogonal SOPs to ensure its security [15]. In the standard BB84 protocol, Alice sends single photons randomly polarized in either HV or QR base. After a long-distance transmission in fiber, those nonorthogonal SOPs are changed into arbitrary polarizations $P_H$, $P_V$, $P_Q$ or $P_R$, respectively. One may apply proper piezoelectronic voltages on $X_1$ and $X_2$ to rotate $P_H$ ($P_V$) back to H(V), or $P_Q$ ($P_R$) back to Q(R). However, there exist no voltages on $X_1$ and $X_2$ that can guarantee simultaneous control of both $P_H$ ($P_V$) and $P_Q$ ($P_R$) back to their original SOPs. This problem can be fortunately solved by using the protocol as shown in Fig. 1(a). At Bob's sites, we added monitoring detectors to collect feedback signals to control the SOPs in the HV- and QR-bases, respectively. As Bob tries to decode polarization information, he actually randomly chooses a HV or QR base to measure the SOPs. Once the decoding base is selected, Bob then only focuses on feedback control of orthogonal polarizations in the selected base, disregarding polarization changes induced in the other bases since only orthogonal polarizations in the selected base give useful decoding. The polarization decoding base can be selected with a 50/50 beam splitter followed by polarization measurement in HV and QR bases, respectively. After the beam splitter, Bob sets feedback signal monitoring, polarization controlling, and polarization decoding in HV and QR bases without cross-interference from the non-orthogonal polarizations. Repetitive feedback control cycles are interrupted within QKD cycles with proper durations to ensure sufficient polarization stability. All the H, V, Q and R detectors are used for monitoring the SOPs in the feedback control cycles, while only clicks on H/V or Q/R detectors are used for HV-based or QR-based polarization decoding in the QKD cycles, respectively. Note that there are 50/50 beam splitters between H/V and Q/R detectors in both HV and QR bases. This is necessary for monitoring SOPs of arriving photons to get feedback signals in the feedback control cycles. However, half of the communication pulses will be disregarded in the QKD cycles as photons reach the monitoring detectors, and consequently, the key creation efficiency will be affected. This can be in principle solved by using electronic polarization rotators before the polarization beam splitter of the monitoring detectors, which electronically switch the monitoring detectors into decoding ones, i.e., from Q/R detectors to H/V ones in the HV base and from H/V detectors to Q/R ones in the QR base, when the feedback control cycles are switched into the QKD cycles.

Although the above-mentioned experimental setup requires more single-photon detectors than the phase-encoded QKD setup, it offers a promising advantage of very low losses in the whole setup from Alice to the detectors at Bob's site, since the main inserting elements before detectors are electronic polarization controllers, which are commercially available with negligible insertion losses. Negligible losses in Bob's decoding setup are of critic significance in long-distance fiber QKD system as absolute security is concerned. As Alice uses no lossy modulation elements, the polarization-encoded QKD is promisingly applicable to ideal single-photon sources. Moreover, polarization scrambling creates additional difficulty for any photon-number-splitting attacks in the quantum channel.

For the feedback control in HV and QR bases, Alice should send H(V) and Q(R) polarizations as reference SOPs, respectively. In the feedback control cycles, Bob's decoding part functions actually as a polarimeter at the single-photon level. With sufficient clicks, the signals of four detectors can form two Stokes parameters $S_1$ and $S_2$. For a typical elliptic polarization state at Bob's site, their normalized values are given by

$$S_1 = \frac{I(H) - I(V)}{I(H) + I(V)} = \cos 2\varepsilon \cos 2\theta,$$
$$S_2 = \frac{I(Q) - I(R)}{I(Q) + I(R)} = \cos 2\varepsilon \sin 2\theta, \quad (1)$$

where I(H), I(V), I(Q) and I(R) are clicks of the H, V, Q and R detectors, respectively, $2\epsilon$ and $2\theta$ are the longitude and latitude of a point on the Poincare sphere, which represent the double azimuth and elliptcity angles, respectively. Due to randomly induced birefringence in fiber, linearly polarized pulses sent by Alice as a reference change to be elliptically polarized with random $\epsilon$ and $\theta$ at Bob's site.

As an experimental demonstration, we set the reference SOP at the horizontal direction (H) and show how single-photon polarization can be stabilized in a long-distance fiber. The schematic of the experimental configuration is shown in Fig. 1(b). The whole communication system was synchronized by repetitive pulses from the laser diode $LD_0$. A separate fiber channel was used for the synchronous pulses in order to avoid influence of intense clock laser and its Raman scattering [16,17]. Pulses from two laser diodes $LD_1$ and $LD_2$ were polarized at vertical and horizontal directions, respectively. Pulses from $LD_1$ were attenuated to 0.1 photons per pulse after $Attn_3$. There were two paths for $LD_2$ switched by an optical switcher (OSW). One was from $Attn_1$ to $Attn_3$, which was attenuated to 0.1 photons per pulse. The other was from $Attn_2$ to $Attn_3$, which was attenuated to several photons per pulse as reference pulses to monitor SOP. $LD_0 \sim LD_2$ and OSW were controlled by a computer of Alice through a PCI card (NI6251). At Bob's part, the SOPs of photons were adjusted by an electronic polarization controller (EPC, PolaRITE III,

General Photonics) to compensate for the random birefringence in the long-distance fiber. The H, V, Q and R SOP photons were detected by $D_1$, $D_0$, $D_2$ and $D_3$, respectively. Bob used a PCI card to collect signals of four single-photon detectors, and generate two analog signals $V(X_1)$ and $V(X_2)$ to the drivers of the polarization controllers $X_1$ and $X_2$, respectively.

The random variation of SOP depends on the fiber distance and environment. For a typical situation, the SOP can maintain relatively stable within a few minutes in 50 km fiber. We set a counter which periodically interrupted communication and switched to adjust the polarization. Note that the feedback signals are divergent clicks from single-photon detectors, which exhibited unavoidable fluctuations due to the polarization scrambling in fiber and dark-noise clicks of single-photon detection. The click fluctuation was estimated as ±3% in our experimental situation. To minimize influence from such click fluctuations, we operated the system with a repetition rate of 1 MHz and set an accumulation time of one second ($10^6$ counts) for each sampling signal used for feedback control. A feedback-control program for SOP stabilization works as follows. At first, Bob sends a polarization-control ask to Alice through Ethernet and waits for sampling the SOP signals. After receiving the ask, Alice sends reference photons of H-polarization instead of random H/V or Q/R SOP in the QKD cycles. After sampling and calculating $S_1$ and $S_2$, the program judges the SOP change by comparing $S_1$ and $S_2$ with preset thresholds $T_1$ (near 1) and $T_2$ (near 0), respectively, i.e., if $S_1 > T_1$ and $S_2 < T_2$, the program approximates the SOP at Bob's site as H-polarization. Only if Bob has approximately H-polarization does the program send order to stop the feedback control cycles and restart the QKD cycles. If not, the program operates independently the piezoelectronic actuators $X_1$ and $X_2$ in accordance with the measured values of $S_1$ and $S_2$, respectively. $X_2$ is relatively simple to control because $S_2$ varies monotonously with the applied voltage on $X_2$ near the H-polarization. Only until the measured $S_2$ satisfies $|S_2| < T_2$ does the control voltage on $X_2$ remain unchanged, otherwise, the control voltage increases proportionally to the difference between the measured $S_2$ and target value $S_2=0$. This operation makes the SOP rotate along the HV axis in the Poincare sphere, and its repetitive cycles eventually cause θ convergent to 0. $X_1$ is a little difficult to control since its applied voltage is not a monotonous variety of $S_1$ near the H-polarization. The piezoelectronic stress on $X_1$ corresponds to SOP rotation along the QR axis in the Poincare sphere. In order to reach the target $S_1 > T_1$, multiple thresholds are employed in the program to determine whether an increase or decrease is needed for the applied voltage on $X_1$. If $S_1<T_1$, the applied voltage is at first tested with an increase proportional to the difference between the measured $S_1$ and the target value $S_1=1$, and then $S_1$ is compared with an additional threshold $T_3$ ($T_1>T_3$). Once $S_1$ becomes less than $T_3$, the voltage increase/decrease is reversed. The applied voltage on $X_1$ then changes (with determined increase or decrease) proportionally to the difference between the measured $S_1$ and the target value $S_1=1$, and the process repeats until $S_1$ exceeds $T_1$.

We experimentally tested the feasibility and long-term stability of the above-mentioned polarization feedback control in 50, 75 and 100 km fiber systems, with separate coiled fibers of 25 km in each roll. For different fiber lengths between Alice and Bob, we intentionally adjusted the 1-MHz reference pulses to have approximately the same photon counts I(H) +I(V) at Bob's site. The mean photon numbers at Alice's site were about 0.5, 1.6 and 5.1 per pulse for the 50, 75 and 100 km systems, respectively, corresponding to I(H) +I(V)~3200/s at Bob's site for all three cases. In the experiments, the single-photon polarization control was limited by the difference polarization extinction of polarization beam splitters used in the setup. Taking into account the limited polarization extinctions and dark-count probabilities of single-photon detection, we set the $S_1$ and $S_2$ multiple thresholds ($T_1$, $T_2$, $T_3$) as (0.96, 0.05, 0.94), (0.95, 0.08, 0.93), and (0.95, 0.08, 0.93) for the 50, 75 and 100 km fiber systems under test, respectively. And the intervals of periodically interrupted communication were adjusted according to the fiber lengths. Shorter intervals were used for longer fibers. Under our experimental situation, we selected 4.7, 3.1, 1.6 minutes for the 50, 75 and 100 km fiber lengths, respectively. Figure 2 shows typical experimental situation in the 50 and 100 km fiber systems where the controlling voltages $V(X_1)$ and $V(X_2)$ applied on $X_1$ and $X_2$ established stable feedback polarization controls. In order to guarantee a satisfactory polarization stability in the long-term operation, any possible random degenerations of the reference SOP (from H to $P_H$) were regularly tracked with some preset interruption cycles within sufficiently short intervals. With interruption intervals of 4.7 and 1.6 minutes in the 50 and 100 km fiber systems, the measured Stokes parameters $S_1$ and $S_2$ of Bob's polarizations were plotted in Figs. 3(a,c), which tracked SOPs with stable operations of 630 and 400 minutes and interrupted control duration of 92 and 36 minutes, for the 50 and 100 km fiber lengths, respectively. It is clear from the monitored $S_1$ and $S_2$ that SOP at Bob's site was feedback-controlled to Alice's reference polarization H with about 1/10~1/7 of the communication duration interrupted for feedback control. Due to random fluctuation, the control might go beyond controlling at some points as those marked by A, B, C and D in Fig. 3(a,c), the program could automatically adjust the controlling voltages $V(X_1)$ and $V(X_2)$ on $X_1$ and $X_2$ according to the measured feedback signals, and eventually made $S_1$ and $S_2$ return the well-controlled loop. As voltages $V(X_1)$ and $V(X_2)$ applied on both $X_1$ and $X_2$ were accumulated increase or decrease from the previous counterparts in accordance with feedback signals, continuous increase or decrease might cause the voltages to exceed the available ranges of the $X_1$ or $X_2$ drivers (0~150 V). This can be readily avoided by resetting the paranormal controlling voltage with an increase or decrease of a $2\pi$ voltage (52.2 and 49.0 V for $X_1$ and $X_2$, respectively). As shown in Fig. 2, with regular feedback controls, Bob's SOPs typically fluctuated nearby Alice's reference SOPs even at most of the control cycles. With the multiple thresholds $T_{1\sim3}$ selected as above-mentioned, the average $S_1$ and $S_2$ of the controlled SOPs reached as $(S_1,S_2)=(0.96\pm0.01, 0.02\pm0.07)$, $(0.95\pm0.02, 0\pm0.07)$, and $(0.95\pm0.01, -0.03\pm0.07)$ for our tested long-term operations within 630, 587, and 400 minutes for the 50, 75, 100 km fiber systems, respectively. Figure 3 compares the polarization stabilities with and without feedback polarization control in 50, 75, 100 km fiber systems by monitoring at Bob's site the corresponding $S_1$ and $S_2$ of single-photon pulses, from which we can conclude that Bob could actively maintain polarization-stabilized single-photon pulses of the same SOPs with Alice's reference for very long durations. The long-term tests for all the fiber lengths could be in principle operated longer. The current limited factors came from somewhat instable operation of single-photon detectors, which required a little fine adjustment of the coincidence detection after continuous operation about tens hours.

We next check whether such a single-photon polarization control is sufficient for polarization-encoded QKD to provide quantum bit error rates (QBERs) within the absolute security threshold [18]. We used an experimental setup as shown in Fig. 1(b) to distribute quantum keys and checked the stability and QBER of the polarization-encoded QKD in a long-term operation. Alice randomly selected HV or QR base to encode qubits, and the weak pulses were attenuated to 0.1 photons per pulse. Figure 4 shows the experimental tests of the 50, 75 and 100 km fiber systems within a long-term operation of 352, 342 and 189 minutes, giving the average QBERs as $(3.1\pm1.1)$ %, $(4.8\pm1.5)$% and $(6.6\pm2.0)$%, respectively. It is clear that the measured QBERs increased with the fiber length. As the attenuated single-photon pulses reached 0.006, 0.002 and 0.0006 per pulse after the 50, 75 and 100 km fiber transmissions (including 2dB loss from on-line optical elements), respectively, while the dark counts of the single-photon detectors used in the experiments were 4E-7 (detector D0, quantum efficiency ~20%) and 8E-7 (detector D1, quantum efficiency ~20%), the signal to noise ratio unavoidably became worse as the transmission length increased. The dark-count noise produced ~2% QBER in the 100 km system while it was negligible in the 50 km system. In addition, the vertical SOP would degenerate with fiber length, and would change after a long-time operation in the long-distance fiber, although the horizontal SOP could be stabilized with comparable $S_1$ in fiber systems of different fiber lengths. Gradual degeneration of $S_2$ stabilization affected the long-term stability of QKD. For instance, stable polarization encoding could only be maintained for 3 hours in the 100 km system as the corresponding degeneration of $S_2$ became somewhat serious despite that horizontal polarization could be stabilized continuously. It has been pointed out that random polarization changes in long-distance fiber was dependent on the central wavelengths of the carrier pulses [19]. While the laser diodes ($LD_{0~2}$) used in our experiments were DFB lasers without temperature controlling, and accordingly, the laser central wavelengths differed by about 3 nm, which might produce polarization scrambling slightly dependent on the laser diodes, leading to some instable origins for polarization feedback controlling. High-performance photon sources, such as temperature-controlled DFB lasers with narrow bandwidths, would effectively improve stability of the polarization-encoded QKD.

In conclusion, we have achieved polarization feedback control in long-distance fibers at single-photon level, which facilitated polarization-encoded QKDs with long-term stabilities. Experimental test of polarization encoding in 50, 75 and 100 km fibers demonstrated that the single-photon polarization scrambling in long-distance fibers could be controlled to provide QBERs below the absolute security threshold for polarization-encoded "one-way" QKD in fiber. Our experimental tests were based on single-photon polarization stabilization and encoding in the HV base. As a whole, an integrated QKD requires the same implementation of QR polarizations, which could be readily fulfilled with a similar setup in the QR base. Polarization-encoded "one-way" QKD is not only promisingly applicable to ideal single-photons, but also possesses an enhanced security due to its low losses in the whole setup from Alice to the detectors at Bob's site.

This work was funded in part by National Natural Science Fund (Grants 10525416 and 10374028), National Key Project for Basic Research (Grant 2001CB309301), key project sponsored by National Education Ministry of China (Grant 104193), and ECNU PhD Scholarship.

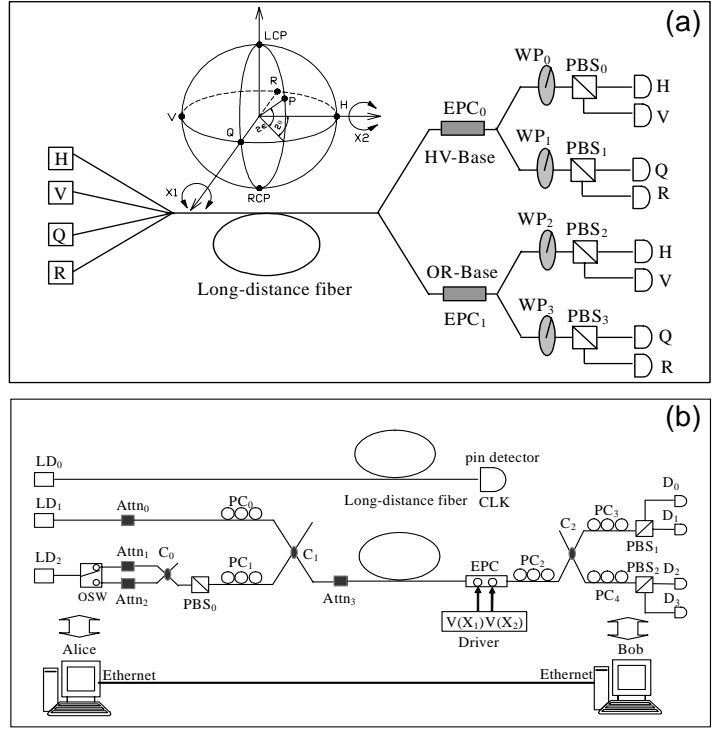

Figure 1 (a) The schematic of single-photon polarization stabilization based on the BB84 protocol, where single-photon detectors detecting H, V, Q and R polarized photons, $EPC_0$ and $EPC_1$ are the electronic polarization controllers corresponding to $X_1$ and $X_2$ that make the SOP rotate around QR and HV axes in the Poincare sphere as shown in the inset, respectively. $WP_{0-3}$: quarter-wave plates, $PBS_{0-3}$: polarization beam splitters. (b) The schematic setup of single-photon polarization stabilization in the HV base. $LD_{0\sim2}$: DFB laser diodes at 1550 nm, $Attn_{0\sim3}$: variable optical attenuators, $PC_{0\sim4}$: fiber polarization controllers, OSW: optical switcher, $C_{0\sim2}$: 50/50 fiber optical splitters, $D_{0\sim3}$: single-photon detectors.

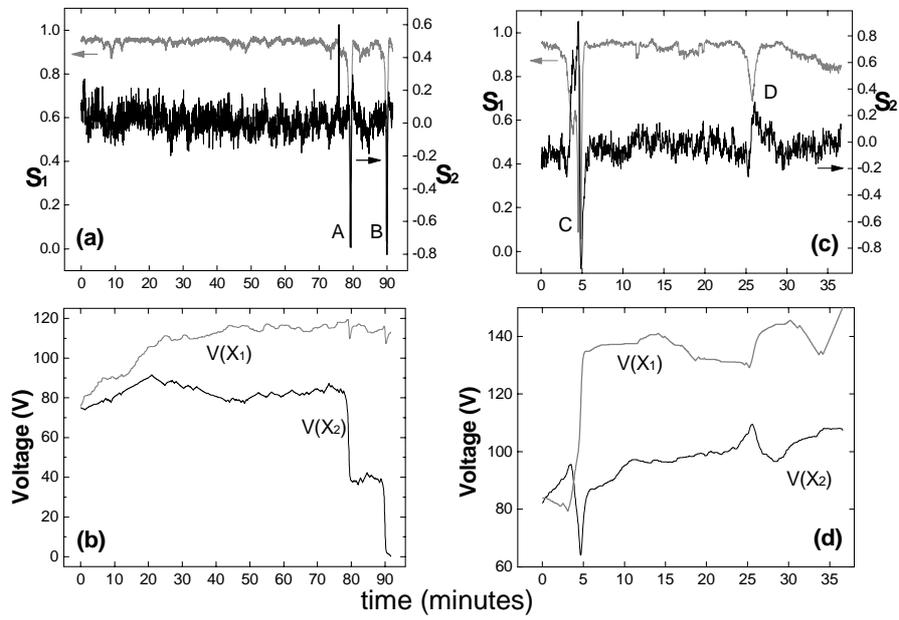

Figure 2 Feedback signals $S_1$ and $S_2$ (a,c) and their corresponding controlling voltages $V(X_1)$ and $V(X_2)$ (b,d) for the 50 km (a, b) and 100 km (c,d) fiber systems.

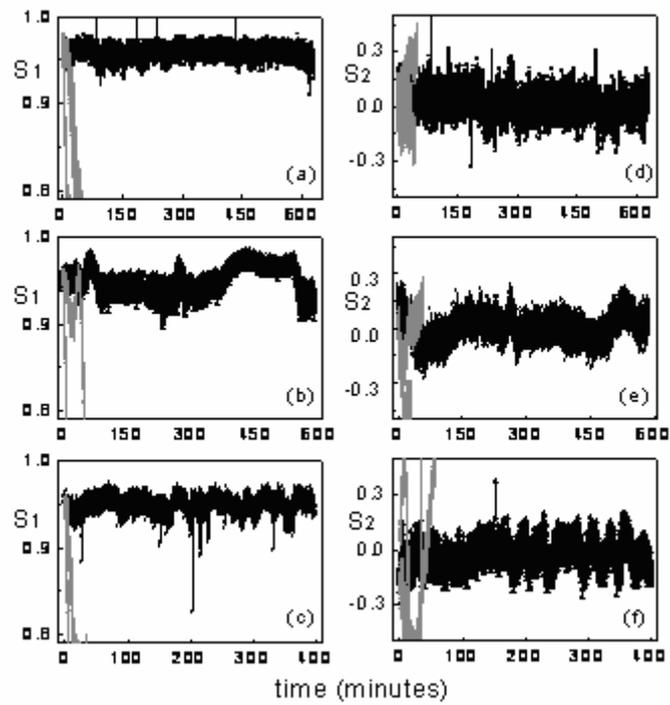

Figure 3 The comparison of the single-photon polarization variation with (dark lines) and without active feedback controls (grey lines) for $S_1$ (a, b, c) and $S_2$ (d, e, f) monitored in long-term operations of 50 km (a,d), 75 km (b, e), and 100 km (c, f) fiber systems.

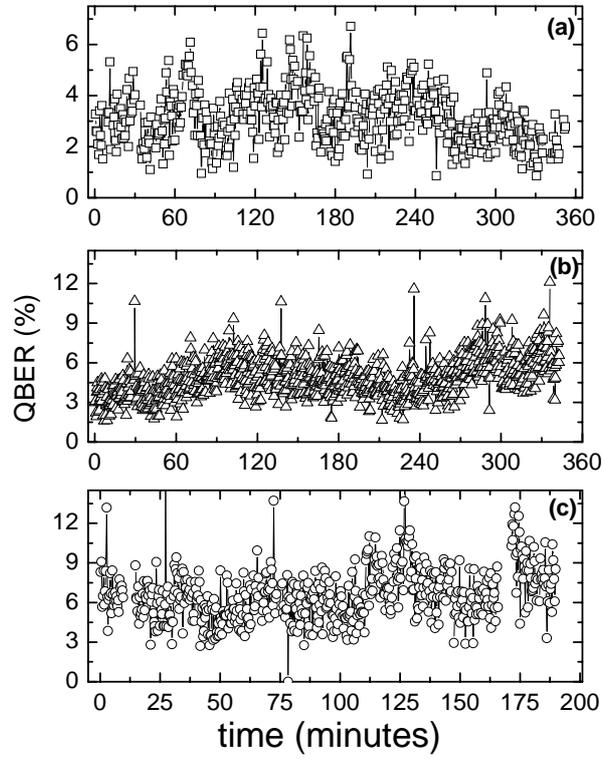

Figure 4 QBERs of polarization-encoded QKD in 50 km (a), 75 km (b), and 100 m (c) fiber system, respectively.